\documentclass{aa}
\usepackage{graphicx, textcomp, color}
\usepackage{natbib}
\usepackage[usenames,dvipsnames]{xcolor}
\usepackage{txfonts}
\usepackage{hyperref}
\hypersetup{
     colorlinks   = true,
     citecolor    = violet
}

\begin{document} 

%\linenumbers
   \title{The stratified two-sided jet of Cygnus A}
   \subtitle{Acceleration and collimation}

   \author{B. Boccardi$^1$
\and T.P. Krichbaum$^1$
\and U. Bach$^1$
\and F. Mertens$^1$
\and E. Ros$^{1,2,3}$
\and W. Alef$^1$
\and J.A. Zensus$^1$
%\inst1
}

          \institute{{$1$: Max-Planck-Institut f\"{u}r Radioastronomie, Auf dem H\"{u}gel 69, D-53121 Bonn, Germany \\ 
        $2$: Observatori Astron\`{o}mic, Universitat de Val\`{e}ncia, C.\ Catedr\'{a}tico Jos\'{e} Beltr\'{a}n 2, E-46980 Paterna, Val\`{e}ncia, Spain\\
        $3$: Departament d'Astronomia i Astrof\'{\i}sica, Universitat de Val\`{e}ncia, C.\ Dr.\ Moliner 50, E-46100 Burjassot, Val\`{e}ncia, Spain \label{inst1}}}

   \date{Received x, 2015/
 accepted x, 2015}

% \abstract{}{}{}{}{} 
% 5 {} token are mandatory
 
  \abstract
   {}
{High-resolution Very-Long-Baseline Interferometry observations of relativistic jets are essential to constrain fundamental parameters of jet formation models. 
At a distance of 249 Mpc, Cygnus A is a unique target for such studies, being the only Fanaroff-Riley Class II radio galaxy for which a detailed sub-parsec scale imaging of the base of both jet and counter-jet can be obtained. Observing at millimeter wavelengths unveils those regions which appear self-absorbed at longer wavelengths and enables an extremely sharp view towards the nucleus to be obtained.}
   {We performed 7 mm Global VLBI observations, achieving ultra-high resolution imaging on scales down to 90 $\mu$as. This resolution corresponds to a linear scale of only ${\sim}$400 Schwarzschild radii (for $M_{\mathrm{BH}}=2.5 \times 10^9 M_{\odot}$). We studied the kinematic properties of the main emission features of the two-sided flow and probed its transverse structure through a pixel-based analysis.}
   {We suggest that a fast and a slow layer, with different acceleration gradients, exist in the flow. The extension of the acceleration region is large (${\sim} 10^4 R_{\mathrm{S}}$), indicating that the jet is magnetically-driven. The limb brightening of both jet and counter-jet and their large opening angles ($\phi_\mathrm{J}{\sim} 10^{\circ}$) strongly favor a spine-sheath structure. In the acceleration zone, the flow has a parabolic shape ($r\propto z^{0.55\pm 0.07}$). The acceleration gradients and the collimation profile are consistent with the expectations for a jet in ``equilibrium'' \citep{2009ApJ...698.1570L}, achieved in the presence of a mild gradient of the external pressure ($p\propto z^{-k}, k\leq2$).
    }
   {}


   \maketitle

%
%________________________________________________________________
\section{Introduction}
The physics of extra-galactic relativistic jets is one of the most exciting and challenging topics in astrophysics. Considerable progress has been made in its understanding over the last decades. By interpreting jets as fluid phenomena, their properties on parsec and kilo-parsec scales are quite well explained by applying the laws of relativistic hydrodynamics. However, a fundamental problem is still unsolved: how can jets be accelerated up to Lorentz factors of tens, and how can they be so sharply collimated? The answer to these questions must be intimately related to the mechanism of jet launching. The necessity of having light jets, able to reach relativistic speeds, gradually led to the exclusion of a purely hydrodynamic launching, i.e. exploiting the gas pressure gradient in the accretion disk, as this implies too high a mass loading. The most successful models are those involving the extraction of rotational energy from the accretion disk \citep{1982MNRAS.199..883B} or of the black hole itself 
\citep{1977MNRAS.179..433B} through the magnetic field. Some numerical simulations \citep {2004ApJ...611..977M, 2005ApJ...620..878D} have suggested that launching from the accretion disk can only produce jets with moderate Lorentz factors ($\Gamma<3$), while a process involving the black hole rotation can more easily achieve $\Gamma{\sim}10{}$ \citep{2006MNRAS.368.1561M}. However it is not ruled out that the inner disk can launch relativistic jets as well \citep{2004ApJ...605..656V,2007MNRAS.380...51K}. The possibility of reaching a certain terminal Lorentz factor depends strictly on the details of the process which converts the initially magnetically dominated jet into a kinetically dominated one. In theoretical models describing the conversion of Poynting flux, the mechanisms of acceleration and collimation of the flow show a strong interplay. The magnetic field structure is, in this context, a fundamental parameter, but certainly not the only one. In jets missing a large scale (poloidal) 
magnetic field, i.e. featuring a purely toroidal field \citep{1996MNRAS.279..389L}, an efficient magnetic self-collimation seems unlikely \citep{1997MNRAS.288..333S}, and confinement from an external medium appears to be the only possibility. The pressure from a poloidal magnetic field, on the other hand, can effectively confine the flow. However, an additional contribution from the external medium is anyhow necessary for preventing the terminal speed and collimation of the flow to be reached on unreasonably large scales \citep{1994PASJ...46..123T,1998MNRAS.299..341B}. In accreting systems like AGN, this medium is likely to be the magnetized slow wind ejected by the edges of the accretion disk. 
%__________________________________________________________________
\begin{table*}[width=\textwidth]
\caption{\footnotesize{Log of observations at 7\,mm (43 GHz) and characteristics of the clean maps. Col. 1: Day of observation. Col. 2: Array. VLBA - Very Long Baseline Array; GBT - Green Bank Telescope; On - Onsala; Nt - Noto; Eb - Effelsberg; Ys - Yebes. Col. 3: Correlator. Col. 4: Beam FWHM and position angle. Col. 5: Peak flux density. Col. 6: Noise. All values are for untapered data with uniform weighting.}}
\centering
\begin{tabular}{c c c c c c}
\hline
\hline
 Date & Antennas & Correlator & Beam FWHM $\mathrm {[mas, deg]}$& S$_{\mathrm {peak}} \mathrm {[mJy/beam]}$& rms $\mathrm {[mJy/beam]}$  \\
\hline
23/10/2007 & VLBA*, GBT, On, Eb, Ys &JIVE MKIV&$0.23\times 0.11, -21.9^{\circ}$&197&0.13 \\
16/10/2008 & VLBA*, GBT, On, Nt, Eb, Ys &JIVE MKIV&$0.22\times 0.10, -11.0^{\circ}$&264&0.11  \\ 
19/03/2009 & VLBA, GBT, On, Nt, Eb, Ys &Bonn MKIV&$0.23\times 0.09, -12.5^{\circ}$&269&0.14  \\
11/11/2009 & VLBA, GBT, On, Nt, Eb     &Bonn MKIV&$0.19\times 0.10, -20.1^{\circ}$&226&0.32  \\ 
\hline
\end{tabular}\\
\hspace{-10cm}\footnotesize{*Saint Croix data not available.}
\end{table*}

Exact analytic solutions for these problems are difficult to obtain if one takes adequately into account the high degree of complexity of the system. While numerical simulations can considerably improve our knowledge, direct observations of the fundamental parameters of the flow on small scales are equally decisive. 
With the development of VLBI techniques at millimeter wavelengths and of space-VLBI, the spatial resolution achievable in observations of radio loud AGN has dramatically improved over the last few years, allowing us to zoom into the innermost regions of extra-galactic jets. 
Besides the possibility of obtaining explicit measurements of speeds, scale of the acceleration and shape of the flow, high resolution VLBI observations also allow the internal structure of the flow to be examined. Recent parsec and sub-parsec scale imaging has revealed that relativistic jets are far from the simple assumption of homogeneous conical outflows, and show instead a complicated transverse structure. Among other characteristics, limb brightening was observed both in some nearby FRI radiogalaxies, like M\,87  \citep{2007ApJ...668L..27K} and 3C\,84 \citep{2014ApJ...785...53N}, and in blazars, as in the case of Mrk 501 \citep {2004ApJ...600..127G} and 3C\,273 \citep{2001Sci...294..128L}.
Limb brightening has often been interpreted as the result of a spine-sheath structure of the jet, i.e. made of a fast and light ultra-relativistic spine embedded in a cylindrical flow of slower and denser material \citep{1989MNRAS.237..411S}. Such a scenario provides a good explanation for some inconsistencies between observations and the standard unified model for AGN \citep{1989ApJ...336..606B}, as for example the lower Lorentz factors and lower intrinsic jet powers derived in FRI galaxies (sheath-dominated) with respect to BL Lac objects (spine-dominated) \citep{2009A&A...494..527H, 2000AJ....120.2950X}. Also, a spine-sheath interaction seems to be an essential ingredient for Spectral Energy Distribution (SED) modeling of TeV BL Lacs, where the high energy emission can hardly be reconciled with the low Lorentz factors deduced from VLBI studies \citep{2005A&A...432..401G}. The origin of such stratification is however unclear. A two-component jet with the described properties could directly arise \citep{
2007Ap&SS.311..281H, 2012RAA....12..817X} from a jet launching process in which the external sheath is ejected from the accretion disk, while the central spine is fueled from the ergosphere. Alternatively, as shown in \cite{2001Sci...294..128L} for 3C\,273, a double-rail structure can also be effectively created by helical Kelvin-Helmholtz instabilities. In this case, if the emission is dominated by the pressure-enhanced regions of the instability, the observed motions will be those of the instability patterns, typically propagating at much lower speeds (0.1$-$0.5 c) compared to the flow or the shocks in the jet. 

In the following we present a study of the jet transverse structure and kinematics of the Fanaroff-Ryley II (FRII) radio galaxy Cygnus A, from Global VLBI observations at 43 GHz. At the source redshift (z=0.056, 1 mas${\sim}$1.084 pc, assuming a $\Lambda$CDM cosmology with H$_\mathrm{0}$= 70.5 h$^{-1}$ km s $^{-1}$ Mpc $^{-1}$, $\Omega_\mathrm{M}=0.27$, $\Omega_{\mathrm{\Lambda}}=0.73$), these provide an angular resolution of ${\sim}$400 Schwarzschild radii for $M_{\mathrm{BH}}=2.5 \times 10^9 M_{\odot}$  \citep{2003MNRAS.342..861T}, and allow the immediate surroundings of the central engine to be imaged. In addition to the proximity, the choice of this target is motivated by further advantages. Firstly, the jet of Cygnus A is seen at a large viewing angle, in the range $50^{\circ}-85^{\circ}$ \citep[and references therein]{1995PNAS...9211371B}. This means that both geometrical and relativistic effects are much reduced, implying respectively that the intrinsic morphology and speed of the flow can 
be more easily inferred and that a counter-jet can be detected. Moreover, both the main jet and the counter-jet can be transversally resolved on sub-parsec scales, which is fundamental for studying the collimation. Ultimately, while more such studies have been carried out on FRI radiogalaxies (e.g M87, 3C84, CenA), few observational constraints have been provided for strong FR II objects, mainly because their jets are generally fainter on parsec scales. Cygnus A is in this respect a unique target. The paper is organized as follows.  In Sec. 2 we present the observational setup and data reduction; in Sec. 3 we describe the methods used for the kinematic analysis and the study of the transverse intensity profiles; in Sec. 4 \& 5 we present and discuss the results, which are summarized in Sec. 6.
\section{Observations and data reduction}
The data set comprises 4 epochs at 7\,mm (43\,GHz) from Global VLBI observations made between 2007 and 2009 (Tab. 1). The long total on-source time of ${\sim}$ 8 hours, spanned over a ${\sim}$16 hour track, and the large number of antennas (varying between 13 and 15) produced a good uv-coverage. 
During observations, antennas were frequently switched between the target and the nearby calibrator 2013+370, in order to reduce pointing errors at the biggest dishes. 
For the first three epochs, data were recorded in dual polarization mode using 4 subbands (IFs) with a total bandwidth of 64 MHz per polarization (512 Mb/s recording rate). The last epoch, instead, was observed in single polarization, with 16 subbands and a total bandwidth of 128 MHz. Data were calibrated in AIPS (Astronomical Image Processing System) following the standard procedures, including fringe fitting, opacity correction for atmospheric attenuation and bandpass calibration. The imaging and self-calibration of amplitude and phase were performed in DIFMAP  (Difference Mapping).   
\section{Data analysis}
\subsection{Model fitting and alignment of the images}
Since the absolute position information is lost during phase self-calibration of VLBI data, one of
the main difficulties in multi-epoch studies is image registration. This can be particularly challenging in the case of radio galaxies because, unlike in blazars, a prominent VLBI core to be taken as the stationary reference point is usually missing.
In order to obtain a simplified model of the source, we have fitted circular Gaussian components to the complex visibilities using the  \textsc{MODELFIT} subroutine in  \textsc{DIFMAP}. The components were then cross-identified in the 4 epochs, taking into account their size, flux density and the morphology of the region (Fig. 1 \& Tab. 2). The images were finally shifted and aligned on the position of component N, assumed to be stationary. This feature, although it may not coincide with the true jet base, is the most compact and describes a region with negligible morphological changes. The same choice was made in \cite{1998A&A...329..873K} (there named J0), based also on a spectral analysis. Note that, by applying the aforementioned shift, the positions of a sharp gap of emission at $\sim$ 0.2 mas to the east of component N also become aligned. The nature of this feature, already noticed in 
\cite{2008evn..confE.108B} is discussed in Sec. 5.6.
According to the cross-identification described, a new component (J1) was ejected between October 2007 and March 2009, accompanied by a brightening by about 50\% of the core and inner-jet region. The kinematics resulting from this alignment will be presented in Sect. 4.1. 

Concerning the errors associated with the each parameter of the gaussian components, reported in Tab. 2, a rigorous method taking into account the local signal to noise in the image around each feature was applied \citep[see][for the details of the method]{2005astro.ph..3225L, 2012A&A...537A..70S}. Since the positional errors obtained are quite small compared to the beam size, conservative values equal to $10\sigma$ for the radial separation (Col. 3) and $5\sigma$ for the transverse size (Col. 5) were assumed.
\begin{figure}[!ht]
\centering
\includegraphics[trim=0cm 4.8cm 7.5cm 0cm, clip=true, scale=0.2]{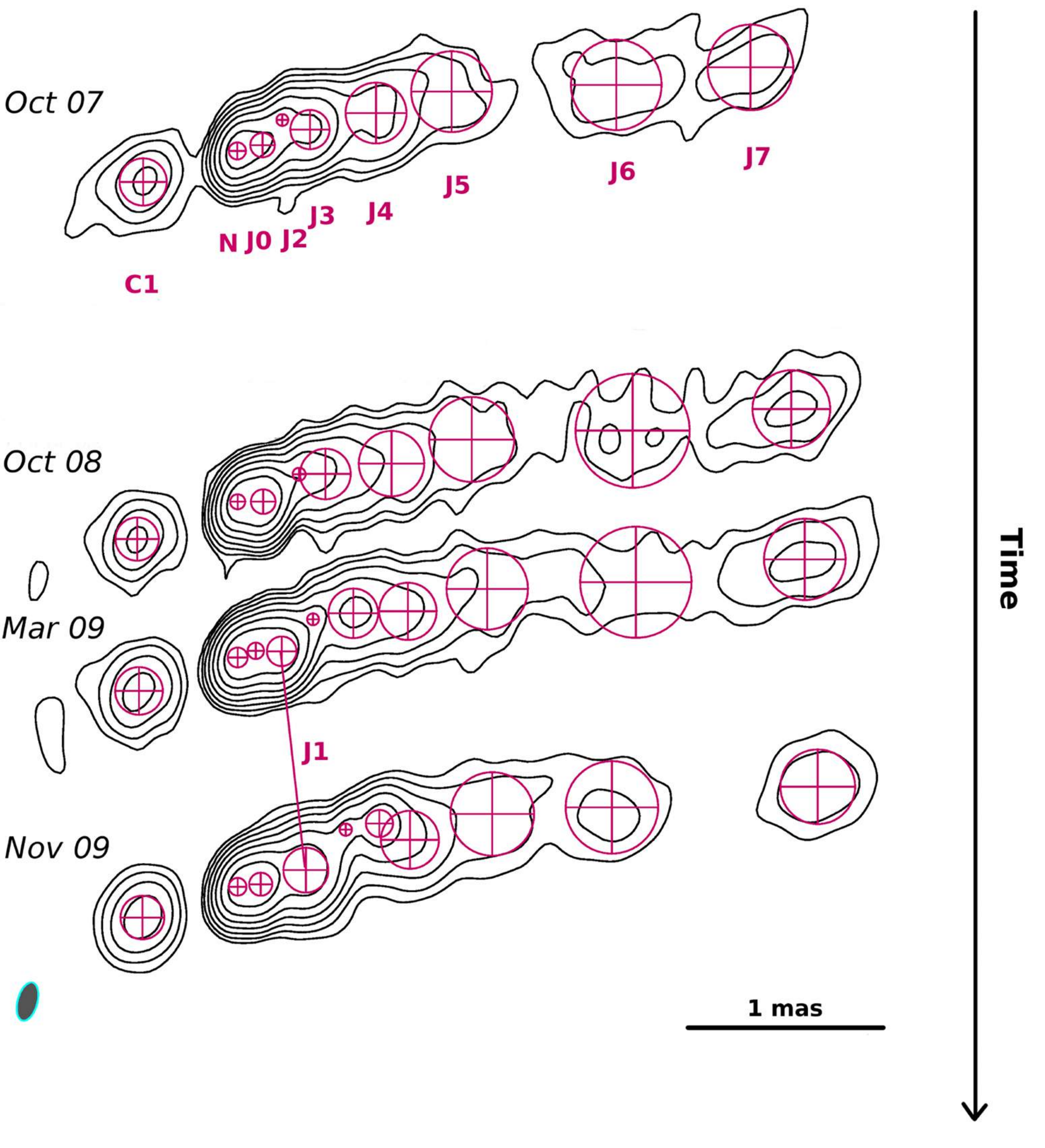}
\caption{\footnotesize{43 GHz modelfit maps of CygA. Images were tapered and convolved with a common beam with FWHM of 0.2 $\times$ 0.1 mas, \textit{pa}=-15$^{\circ}$. Contours represent isophotes at 1.8, 3.6, 7.2, 14.4, 28.8, 57.6, 115.2 mJy/beam. The reference point for the alignment is component N, assumed to be stationary. }}
\end{figure}
\subsection{Transverse structure}
For analyzing the transverse structure of the jets in Cygnus A, the four \textsc{CLEAN} images were convolved with a common circular beam 
of 0.15 mas, which is approximately intermediate between the major and minor axes of the natural beams
(Tab. 1). Therefore, a slight super-resolution in the transverse direction was applied. The analysis was performed both on the single epoch maps and on a stacked image (Fig. 2), created after aligning the four maps as described in Sect. 3.1. In order to facilitate the study, the images were first rotated by 16 degrees clockwise, so that the mean jet axis was parallel to the $x$ axis, and sliced pixel-by-pixel (1 px=0.03 mas) using the \textsc{AIPS} task \textsc{SLICE} in the transverse direction. Then, the transverse intensity distributions of each slice were fitted with Gaussian profiles using the task \textsc{SLFIT}. This task was ran twice for the entire length of the jets, first fitting a single Gaussian and then fitting two Gaussian profiles. 
\begin{table}[!h]
\caption{\footnotesize{Properties of the \textsc{MODELFIT} gaussian components from the first to the last epoch (1: 2007.81, 2: 2008.79, 3: 2009.21, 4: 2009.86). 
Col. 1: Component label. Col. 2: Epoch. Col. 3: Radial separation from the core N. Col. 4: Integrated flux density. Col. 5: Transverse size (FWHM). Col. 6: Apparent speed in units of $c$.}}
\centering
\footnotesize
\begin{tabular}{c c c c c c}
\hline
\hline
ID   &Ep.&  $z~\mathrm{[mas]}$ & $S_{\mathrm{\nu}}~\mathrm{[Jy]}$ & $r~\mathrm{[mas]}$ & $\beta_{\mathrm {app}}$ \\
   \hline
C1 &1& -0.504$\pm$0.043           & 68$\pm$14            &  0.240$\pm$0.043&                         \\
   &2& -0.545$\pm$0.063           & 61$\pm$16            &  0.220$\pm$0.063&                         \\
   &3& -0.529$\pm$0.041           & 78$\pm$15            &  0.243$\pm$0.041&                         \\
   &4& -0.508$\pm$0.046           & 79$\pm$17            &  0.222$\pm$0.046&    0.00$\pm$0.06        \\
   \hline                                                                
N  &1&   0                       & 254$\pm$20           &  0.087$\pm$0.002&                         \\
   &2&   0                       & 223$\pm$24           &  0.076$\pm$0.003&                         \\
   &3&   0                       & 313$\pm$31           &  0.101$\pm$0.003&                         \\
   &4&   0                       & 212$\pm$14           &  0.090$\pm$0.002&                         \\
   \hline                                                                     
J0 &1& 0.131$\pm$0.002           & 212$\pm$15           &  0.124$\pm$0.002&                         \\
   &2& 0.130$\pm$0.003           & 473$\pm$39           &  0.125$\pm$0.003&                         \\
   &3& 0.098$\pm$0.004           & 201$\pm$25           &  0.083$\pm$0.004&                         \\
   &4& 0.118$\pm$0.003           & 341$\pm$27           &  0.117$\pm$0.003&     $-$0.03$\pm$0.03    \\
   \hline                                                                
J1 &1& \dots                    & \dots                &  \dots         &                    \\
   &2& \dots                    & \dots                &  \dots         &                    \\
   &3& 0.226$\pm$0.005          & 446$\pm$40           &   0.151$\pm$0.005&                         \\
   &4& 0.357$\pm$0.026          & 359$\pm$55           &   0.228$\pm$0.026&      0.75$\pm$0.04      \\
   \hline                                                                
J2 &1& 0.278$\pm$0.004          & 53$\pm$7             &   0.060$\pm$0.004&                          \\
   &2& 0.342$\pm$0.028          & 44$\pm$11            &   0.066$\pm$0.028&                          \\
   &3& 0.429$\pm$0.039          & 31$\pm$9             &   0.060$\pm$0.039&                          \\
   &4& 0.621$\pm$0.013          & 34$\pm$6             &   0.062$\pm$0.013&     0.96$\pm$0.05        \\
   \hline                                                                 
J3 &1& 0.385$\pm$0.005          & 212$\pm$17           &   0.199$\pm$0.005&                          \\
   &2& 0.467$\pm$0.040          & 143$\pm$27           &   0.257$\pm$0.040&                          \\
   &3& 0.629$\pm$0.026          & 140$\pm$21           &   0.252$\pm$0.026&                          \\
   &4& 0.789$\pm$0.014          & 84$\pm$13            &   0.138$\pm$0.014&     1.15$\pm$0.04        \\
   \hline                                                                  
J4 &1& 0.731$\pm$0.051           & 108$\pm$21           &  0.310$\pm$0.051&                          \\  
   &2& 0.805$\pm$0.121           & 92$\pm$26            &  0.332$\pm$0.121&                          \\
   &3& 0.896$\pm$0.048           & 90$\pm$17            &  0.291$\pm$0.048&                          \\
   &4& 0.907$\pm$0.078           & 94$\pm$22            &  0.395$\pm$0.078&     0.37$\pm$0.07        \\
   \hline
J5 &1& 1.129$\pm$0.113           & 50$\pm$12            &  0.411$\pm$0.113&                          \\ 
   &2& 1.230$\pm$0.245           & 59$\pm$21            &  0.432$\pm$0.245&                          \\
   &3& 1.315$\pm$0.170           & 66$\pm$20            &  0.414$\pm$0.170&                          \\
   &4& 1.345$\pm$0.248           & 64$\pm$23            &  0.425$\pm$0.248&     0.44$\pm$0.05        \\
   \hline
J6 &1& 1.954$\pm$0.171           & 71$\pm$20            &  0.461$\pm$0.171&                           \\
   &2& 2.038$\pm$0.461           & 59$\pm$25            &  0.580$\pm$0.461&                           \\
   &3& 2.058$\pm$0.438           & 64$\pm$27            &  0.566$\pm$0.438&                           \\
   &4& 1.943$\pm$0.209           & 53$\pm$17            &  0.470$\pm$0.209&     0.00$\pm$0.10         \\
   \hline
J7 &1& 2.641$\pm$0.152           & 57$\pm$16            &  0.436$\pm$0.152&                            \\
   &2& 2.851$\pm$0.270           & 71$\pm$20            &  0.395$\pm$0.270&                            \\
   &3& 2.923$\pm$0.127           & 89$\pm$23            &  0.414$\pm$0.127&                            \\
   &4& 2.990$\pm$0.176           & 56$\pm$17            &  0.379$\pm$0.176&      0.67$\pm$0.07         \\
   \hline
\end{tabular}
\end{table}
\begin{figure*}[!ht]
\centering
\includegraphics[width=\textwidth]{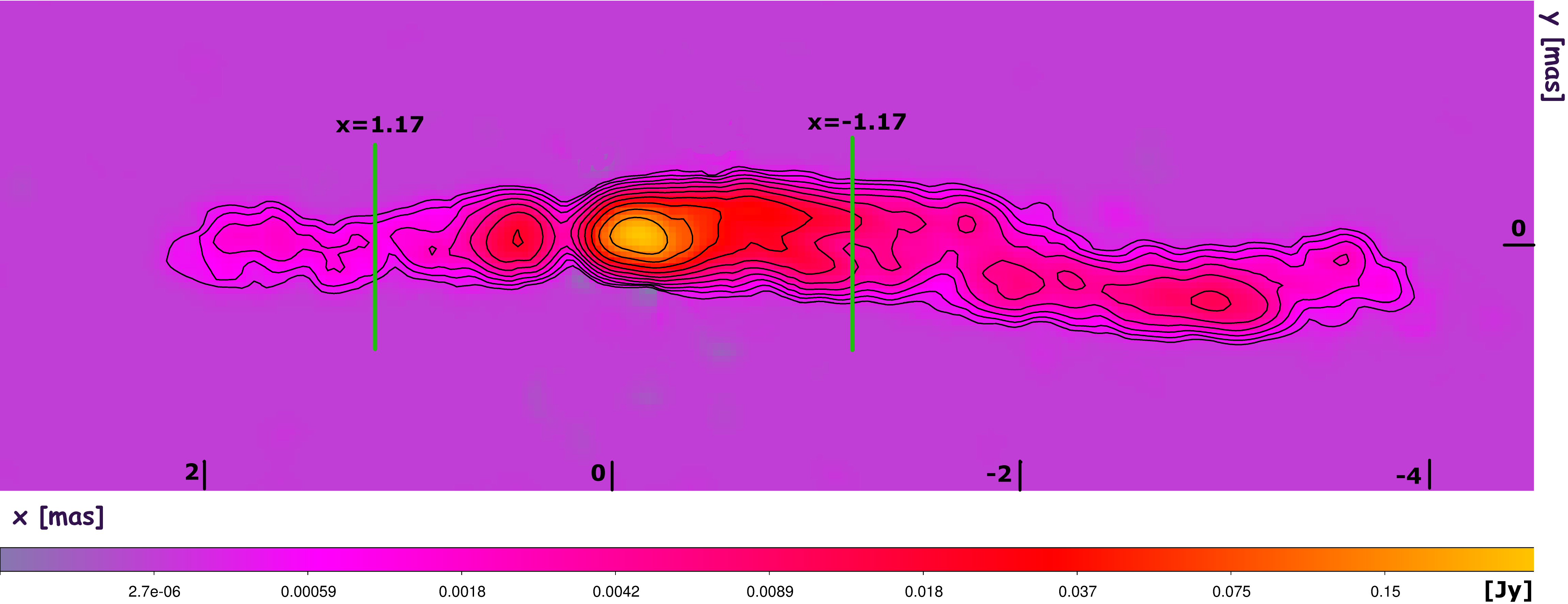}\\
%\hspace{14cm}\tiny{\textbf{[Jy]}}
\includegraphics[scale=0.45]{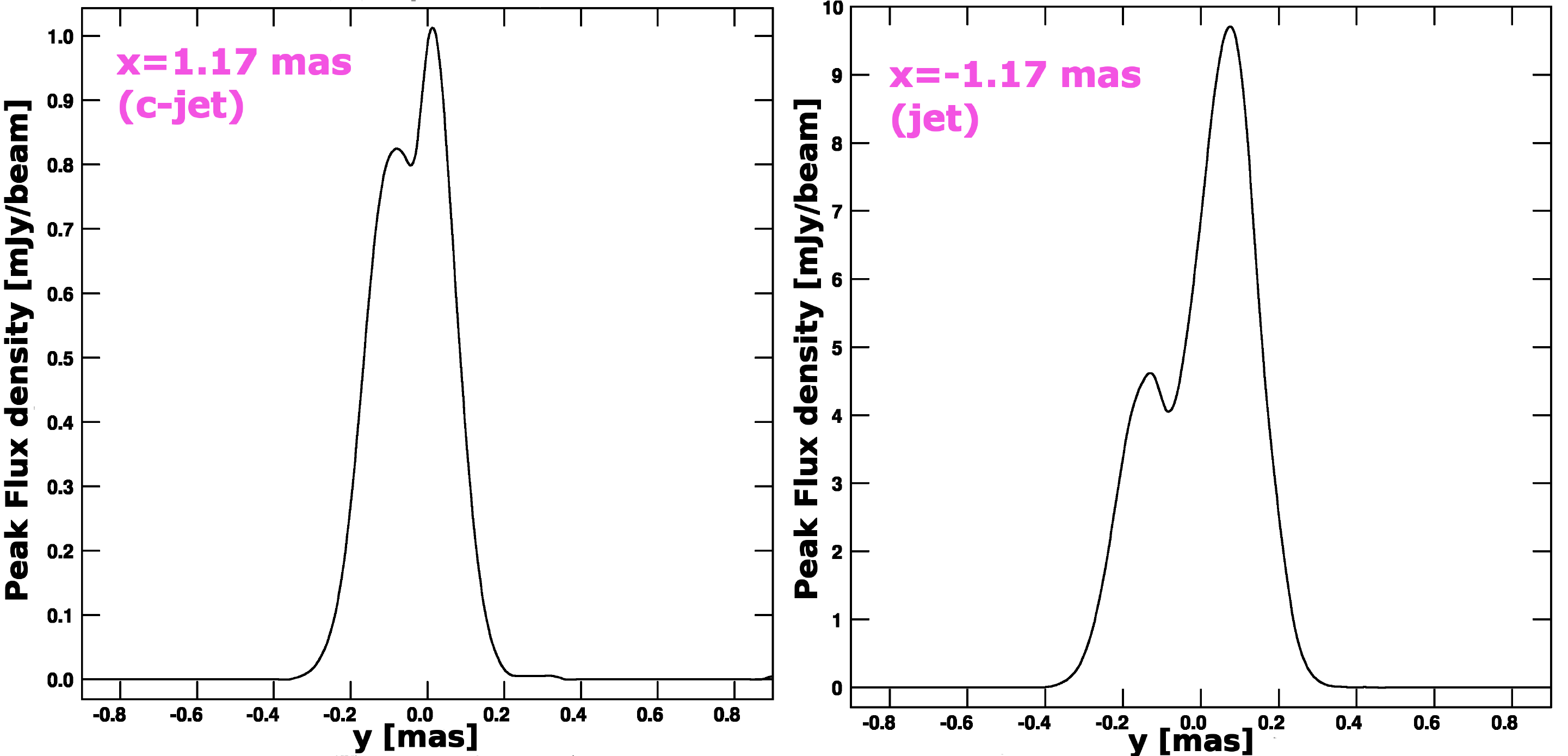}\\
\caption{\footnotesize{\textbf{Top} - Stacked 43 GHz image from observations in 2007-2009, rotated by 16$^{\circ}$ clockwise. It is convolved with a circular beam of 0.15 mas. Contours represent isophotes at 0.3, 0.6, 1.2, 2.4, 4.8, 9.6, 19.2, 38.4, 76.8, 153.6 mJy/beam. The four maps were aligned to the position of component N, as shown in Fig. 1. \textbf{Bottom} - Double peaked intensity profiles at the location of the cuts indicated in the top image. }}
\end{figure*}
\section{Results}
\subsection{Kinematic analysis and components light-curves}
The proper motions $\mu$ of the features described in Fig. 1 were calculated by fitting their radial separations $z$ (Tab. 2, Col. 3) from component N with time $t$, as shown in Fig. 3. The motion of components J2 and J3 is best described by 2nd order polynomials, indicating an acceleration. An estimate of the kinematic parameters was obtained by fitting a simple parabola of the form $z(t)=at^2+c$. The motion of the remaining features, instead, appears more uniform and a linear fit was performed. 
The apparent speed $\beta_{\mathrm {app}}= v_{\mathrm {app}}/c$ can be calculated as
$\beta_{\mathrm {app}} = (\mu D_{\mathrm {L}})/(c(1+Z))$, where $D_{\mathrm {L}}$ the luminosity distance, $c$ is the speed of light and $Z$ is the redshift. In the case of the accelerating features, $v_{\mathrm {app}}$ is the speed calculated for the last epoch, November 2009. The values obtained for the apparent speed of each component are reported in Tab. 2, Col. 6. In the approaching jet we observe an acceleration up to $\beta_{\mathrm {app}} = 1.15 \pm 0.04$ within ${\sim}$ 0.8 mas from the core. Then the speed drops significantly, and increases again in the outer-jet, but remaining largely subluminal. Three components do not show a significant proper motion:  C1 ($\beta_{\mathrm {app}} = 0.00 \pm 0.06$), J0 ($\beta_{\mathrm {app}} = -0.03 \pm 0.03$) and J6 ($\beta_{\mathrm {app}} = 0.00 \pm 0.10$). Given the fact that the entire nuclear region, represented by component N and by the adjacent J0 and C1, appears stationary, a slightly different choice of the reference point for the kinematic study would not 
change the results; therefore the exact core registration is in this case not crucial.
\begin{figure}
\centering
\includegraphics[trim=3cm 0cm 3cm 0cm, clip=true, width=\linewidth]{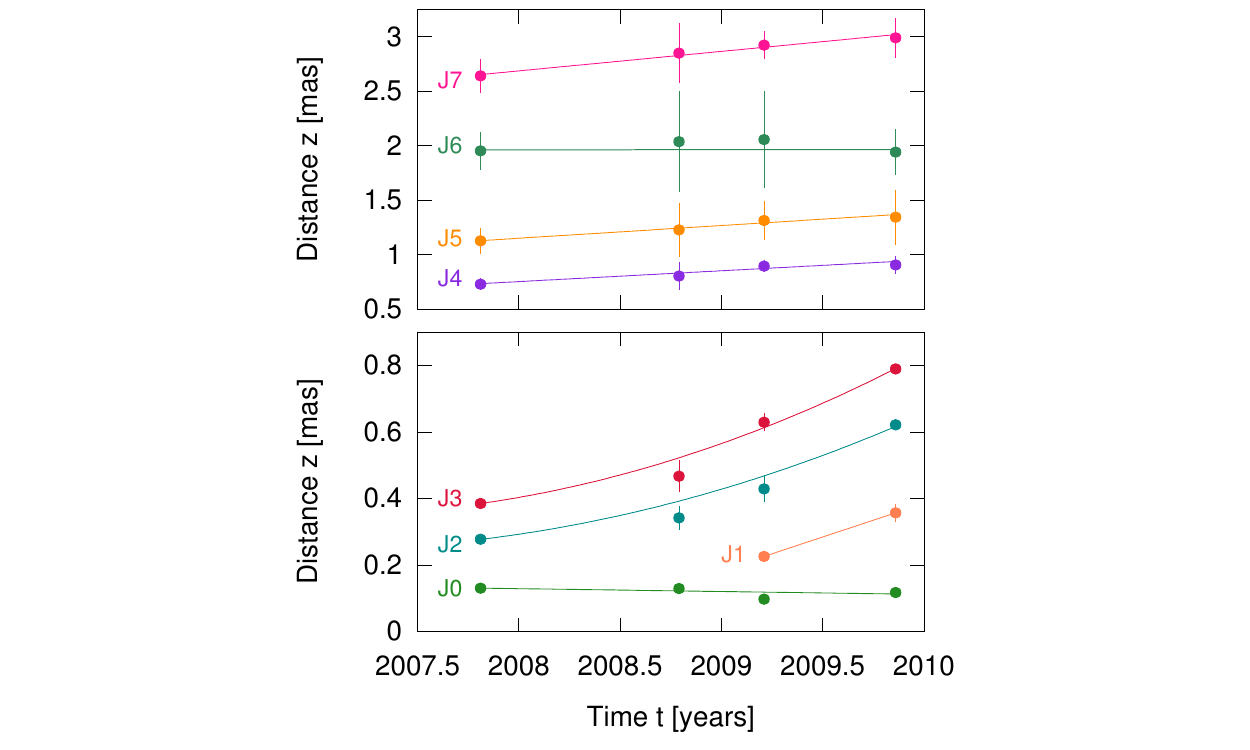}
\caption{Radial distance from core $z$ versus time $t$ for the components in the outer-jet (top) and those in the inner-jet (bottom).}
\end{figure}
The hypothesis that the inner-jet components are crossing an acceleration region is supported by their light-curves (Fig. 4). The flux densities of J1, J2 and J3 are decreasing with time, while those of the outer features (J4-J7) are approximately constant, or increasing in some cases. As the jet expands, a decrease of the energy emitted by the traveling features may be expected because of adiabatic losses. However, we argue that this effect has little impact in our case. The distance traveled by each component during the period of the monitoring is in fact small enough that they do not experience significant changes of the jet width. As shown in Tab. 2 (Col. 5), the sizes of the components are either approximately constant in time or they vary slightly without showing a clear trend.
Instead, differential Doppler (de)-boosting may better explain the observed properties. It is important to keep in mind that the Doppler factor $\delta$ is a monotonically increasing function of speed $\beta$ only for a tiny range of viewing angles, approximately between 0 and 2 degrees. At larger angles, higher speeds often imply a lower boosting, or even a de-boosting ($\delta <1$). The latter means that the narrow boosting cone of the emission can point away from the observer not only in the counter-jet, but also in the approaching side, at a sufficiently high speed exceeding the critical speed $\beta_{\mathrm{\delta=1}}$:
\begin{equation}
\centering
\beta_{\mathrm{\delta=1}} = \dfrac{2 \cos\theta}{1+\cos^2 \theta}
\end{equation}
Flows featuring Lorentz factors as high as 50 are already de-boosted if oriented at an angle of $\geq 11 ^{\circ}$, while a Lorentz factor of ${\sim}$ 1.1 ($\beta {\sim} 0.49$) is sufficient at an angle of $75^{\circ}$.
Since the viewing angle of Cygnus A is certainly large, we find it plausible that the quick and strong decrease in flux density of the inner-jet components is rather due to a decrease of their Doppler factors, which strengthens the scenario suggested by the kinematic analysis.
\begin{figure}[!h]
\centering
\includegraphics[trim=1.3cm 0cm 1.2cm 0cm, clip=true, width=\linewidth]{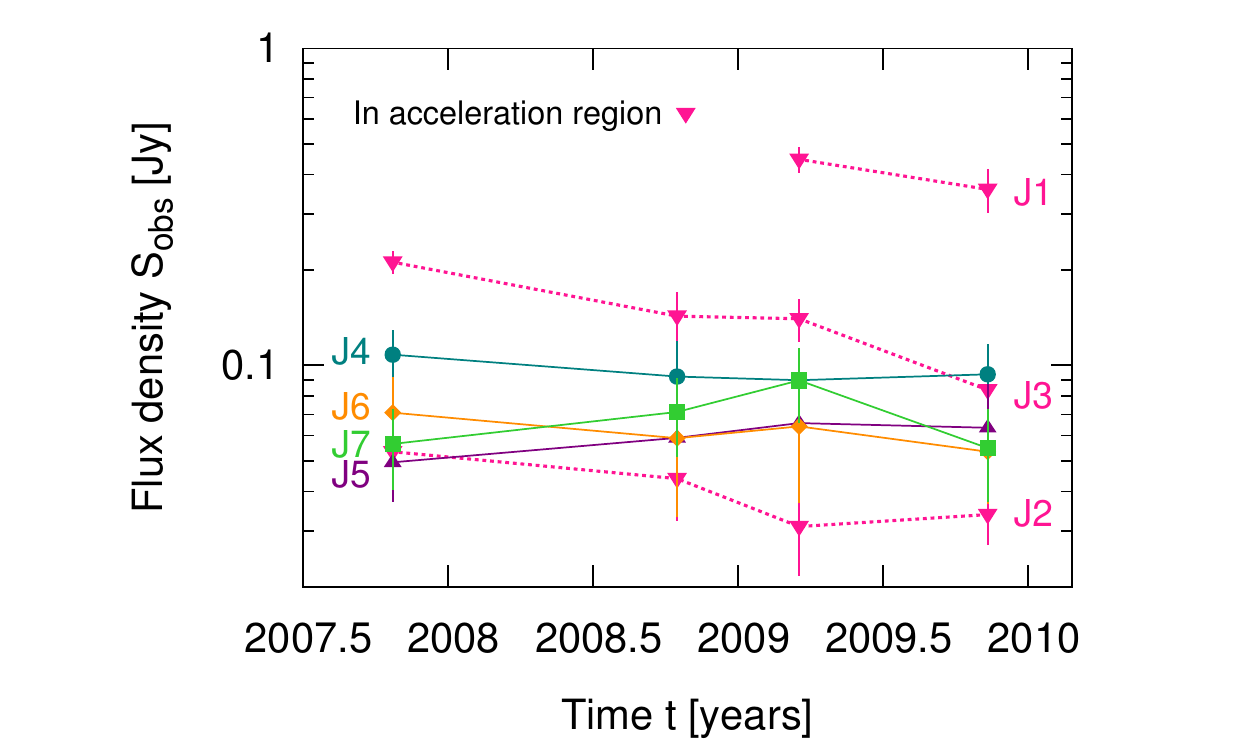}
\caption{\footnotesize{Light-curves of the jet components. The accelerating inner-jet components (dashed-magenta lines) are characterized by a decreasing flux density with time, while the outer ones (continuous lines-other colors), moving at constant speed, feature a more constant flux.}}
\end{figure}
In this framework, the detection of the new feature J1 in March 2009 gives us the opportunity to calculate the Doppler factor of the flow at the position occupied by J1 in November 2009, $z=0.36$ mas, and therefore to obtain an estimate for the jet viewing angle $\theta$. Assuming that the flux density of J1 is close to the intrinsic one in March 2009 (due to the low intrinsic speed we expect at that position) the decrease of flux density 
constrains $\delta$ according to:
\begin{equation}
\centering
\delta = {\left(\frac{S_{\mathrm{obs}}}{S_{\mathrm{int}}}\right)}^{\frac{1}{2-\alpha}},
\end{equation}
where $\alpha$ is the spectral index, $S_{\mathrm{obs}}$ is the flux density of J1 in November 2009 and $S_{\mathrm{int}}$ is that in March 2009. With a classical synchrotron spectral index $\alpha=-0.7$ ($S \propto \nu^{\alpha}$), we obtain $\delta=0.92$. A second constraint on the apparent speed $\beta_{\mathrm{app}}=0.75$ of J1 in November 2009 allows us to solve a system of 2 equations for deriving the viewing angle $\theta$ and the intrinsic speed $\beta$ at $z$=0.36 mas:
\begin{equation}
\centering
\beta_{\mathrm{app}}= \dfrac{\beta \sin \theta}{1-\beta \cos \theta}, ~~~~~~
\delta= \dfrac{\sqrt{1-\beta^2}}{1-\beta \cos \theta}
\end{equation}
obtaining $\beta(z=0.36)=0.65$ and $\theta=74.5^{\circ}$.
The value of the viewing angle obtained can be considered an upper limit, because if the change of Doppler factor is not the only cause for the decrease of flux, a smaller viewing angle is required. 
\subsection{Ridge-line}
The pixel-based analysis of the transverse intensity profiles shows that they are double peaked for most of the length of the source (Fig. 5). The limb brightening is most pronounced in the inner-jet, while a single ridge is visible further out, at a separation from the core between 2 and 4  mas. In the counter-jet we also find evidence for a limb brightened structure: the two ridge lines are closer compared to the main jet, but still well separated after accounting for positional errors, assumed equal to one fifth of each FWHM. An error bar which also takes account of the SNR of the profile \citep[as described in][]{1989ASPC....6..213F} was calculated for comparison. The latter yields unreasonably small errors in the brightest regions, while in the faint ones they are anyhow smaller compared to the first method of calculation. Therefore the most conservative error bars were chosen. In the central ${\sim}$ 0.5 mas, the poor resolution does not allow a double profile to be clearly traced, and results from a 
single Gaussian fit are plotted. The same is done in the other regions of the jet where a well defined double ridge is missing (error bars of the two peaks overlapping or bad fit). The disappearance of the double profile in the main jet (at $z{\sim}$ 2 mas) is accompanied by an apparent bending of the jet axis towards the south and by a narrowing of the jet width, with the two ridges smoothly approaching each other. Interestingly, this position coincides with the location of the stationary component J6. While additional interpretations of this feature are proposed in Sect. 5.3, a more detailed analysis is necessary to establish if the observed structure can be ascribed to helical Kelvin-Helmoltz or to current-driven instabilities developing in the jet. However, following the previous discussion on relativistic effects in flows seen at large viewing angle (in summary: ``the faster, the dimmer''), the most natural explanation for the observed limb brightening is the presence of a transverse gradient of the bulk 
Lorentz factor of a spine-sheath kind, i.e. with $\Gamma$ decreasing from the jet axis towards the edges.
\begin{figure*}[!ht]
\centering
\includegraphics[trim=0cm 1.4cm 0cm 1.4cm, clip=true, scale=1.2]{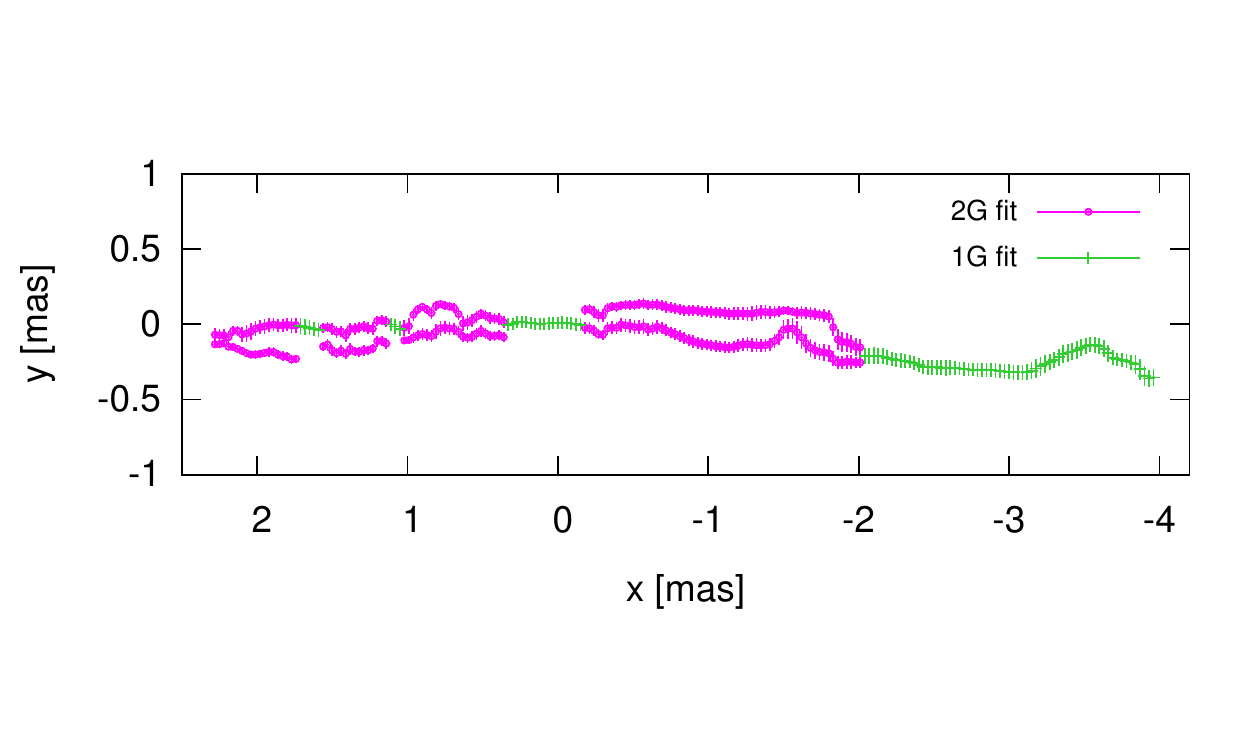}
\caption{\footnotesize{Ridge line structure of Cygnus A at 7 mm. The magenta color is used in those regions were a well defined double ridge line is present, while green points describe the remaining single-peaked regions. The error bars on the peak positions are set to one fifth of the de-convolved FWHM. }}
\end{figure*}
\section{Discussion}
\subsection{Two-layers acceleration}
The kinematic and light-curves analysis of the VLBI maps at 7 mm has revealed the existence of an acceleration region in the inner-jet extending up to a de-projected distance of ${\sim}$0.9 pc, assuming the viewing angle $\theta=74.5^{\circ}$ derived in Sect. 4.1. In the outer-jet, lower speeds are determined, in agreement with findings from VLBI at lower frequencies \citep{2003enig.conf..216B}. The observed fading of the fast flow (Fig. 4) suggests that these lower speeds do not represent an intrinsic deceleration, but are simply the signature that the contribution to the emission from the slower external layers has become dominant in that region. 
If we identify the two regions as representing the ``fast'' and the ``slow'' layers of the flow respectively, we can study separately their kinematic properties. In Fig. 6 (top) we plot the calculated intrinsic speeds $\beta$ versus the de-projected distance from the core expressed in units of Schwarzschild radii $R_{\mathrm{S}}$. We then note that the slow flow is also accelerating, but with a milder gradient compared to the fast one. In this case though, the range of speeds is such that the Doppler factor remains approximately constant, so we do not observe strong variations of the flux density. Both sections are accelerating over a large distance, at least of the order of ${\sim} 10^3 -10^4 R_{\mathrm{S}}$. This is a characteristic signature of relativistic magnetically driven outflows \citep{1994ApJ...426..269B, 1995ApJ...446...67C, 2004ApJ...605..656V}, in which the conversion from Poynting to kinetic flux is found to proceed quite slowly.  The steeper gradient of speed which we measure for the fast flow 
(central spine) compared to the slow one (sheath) indicates that the conversion to kinetic energy proceeds faster for sections of the flow which are closer to the jet axis. By fitting a power law (Fig. 6, top) we find $\beta \propto z^{0.51\pm 0.04}$ for the former, and $\beta \propto z^{0.43\pm0.01}$ for the latter. This translates into a quite different behavior of the Lorentz factor (Fig. 6, bottom), whose dependence on the core separation is $\Gamma \propto z^{0.50\pm0.19}$ and $\Gamma \propto z^{0.11\pm0.04}$ for the two flows respectively.
\begin{figure}[!h]
\centering
\includegraphics[trim=2.5cm 0.2cm 3cm 0.3cm, clip=true, width=\linewidth]{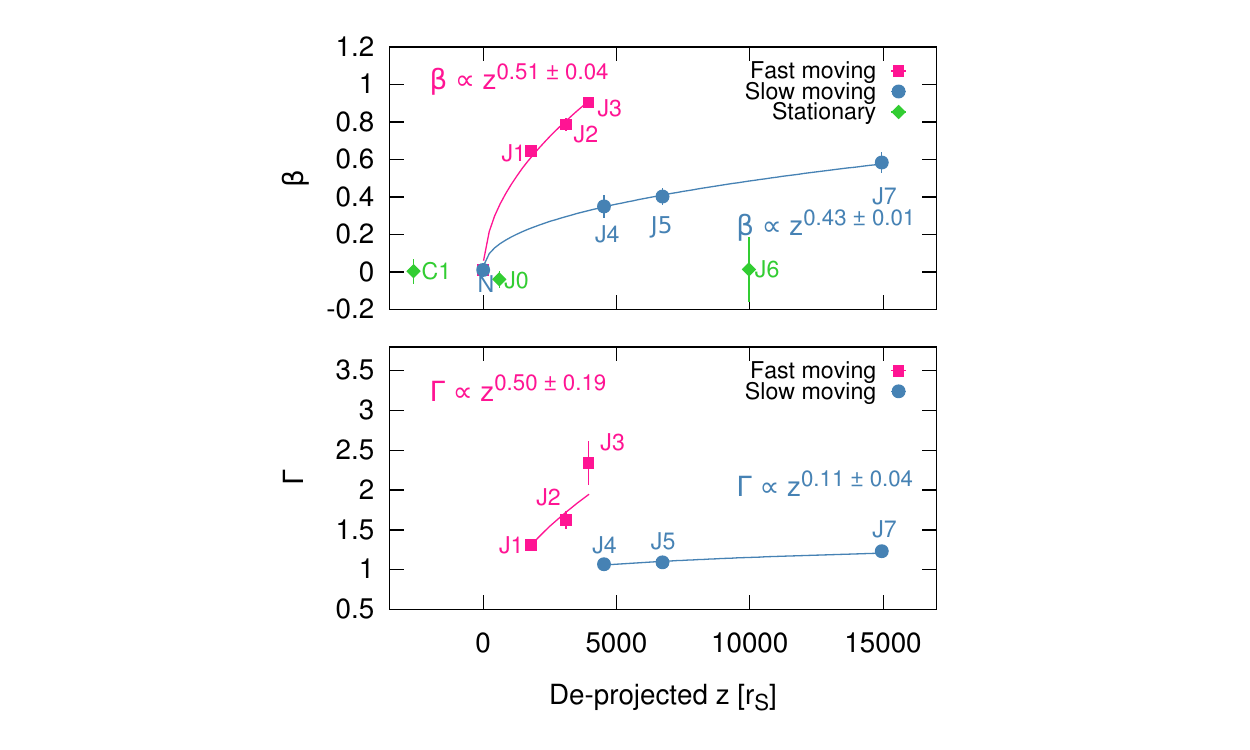}
\caption{\textbf{Top} - Intrinsic speed as a function of de-projected core separation. A fast and a slow component with different acceleration gradients can be identified. \textbf{Bottom} - Corresponding Lorentz factor versus de-projected core-separation.}
\end{figure}
\subsection{Mean opening angle}
From our study, other important parameters of the flow, like the shape and opening angle, can be estimated.
From the analysis of the transverse structure (Sect. 3.2), we can infer the dependence of the jet width on the core separation.
It is preferable in this case to take into account not the stacked map but the single images, as this prevents the temporal evolution of the ridge line from affecting the results. Here we present our findings for the last epoch, November 2009, which is the one showing most prominently the double ridge line structure. The jet width was calculated for each slice as the separation between the edges of the two
(de-convolved) FWHM. The results are shown in Fig. 7, where the de-convolved widths obtained from a single-Gaussian slice fit are plotted for comparison, and show good agreement. This indicates that the expansion of the two ridge lines coincides well with the overall expansion of the jet. We also plot the widths obtained from the modelfit analysis of all the four observing epochs. With some exceptions (discussed below) there is again a substantial overlap. 
The jet width is clearly increasing both in jet and counter-jet up to approximately 2 mas from component N (at $z=0$ in the plot). By performing a linear fit in this region ($|z|<2$) we can calculate the mean apparent full opening angle $\phi_{\mathrm{app}}$, increased relative to the intrinsic one $\phi$ by a factor $1/\sin\theta$ due to projection. In jet and counter-jet we obtain respectively:
\begin{equation}
\phi^{\mathrm{app}}_{\mathrm{j}}=10.2^{\circ} \pm 0.3^{\circ}, ~~~~~~
\phi^{\mathrm{app}}_{\mathrm{cj}}=4.8^{\circ} \pm 0.4^{\circ}
\end{equation}
which for $\theta=74.5^{\circ}$ correspond to intrinsic opening angles:
\begin{equation}
\phi_{\mathrm{j}}=9.8^{\circ} \pm 0.3^{\circ}, ~~~~~~
\phi_{\mathrm{cj}}=4.7^{\circ} \pm 0.4^{\circ}
\end{equation}
As found also in \cite{1998A&A...329..873K} at 22 GHz, the jet appears about two times broader than the counter-jet.
Following the classical assumption of intrinsic symmetry, the apparent opening angle should in principle be the same in jet and counter-jet, unless they are strongly misaligned. Although a small misalignment is present both on pc and kpc scales, this is not enough to justify the observed difference. 
A possible explanation could be the different SNRs of the approaching and receding sides, with the counter-jet being considerably dimmer than the jet at the same core separation and thus detectable over a narrower section. Also, we note that there is, in this case, a certain discrepancy between the modelfit size of C1 and the corresponding width of the intensity profile, with the first being larger and closer to the width of the jet side. This is probably an effect from the smaller beam used in the slicing and it suggests that the counter-jet emission is less compact than the jet emission. However the hypothesis of intrinsic asymmetry, indeed possible \citep[e.g][]{2013ApJ...774...12F}, cannot be excluded.

Based on these results, the jets in Cygnus A can anyhow be defined relatively broad on the examined scale of ${\sim}$2 mas. It is useful to note that the fast jet expansion cannot be clearly perceived by eye in the map in Fig. 2. This is first of all due to the beam convolution effect, causing especially the base of the jet to appear much wider than it actually is. The second reason is that, even after de-convolving the jet width, an apparent full opening angle of ``only'' ${\sim}10^{\circ}$ is difficult to appreciate visually on the scale of the image. 
Despite the misleading appearance, however, the jets in Cygnus A are considerably broader than in blazars, in which intrinsic opening angles of ${\sim}(1-2)^{\circ}$ are typically measured \citep{2009A&A...507L..33P}.
This result, also found by examining other radio galaxies on parsec-scale, represents another piece of evidence for the spine-sheath scenario, in which sources seen at larger viewing angles become more and more sheath-dominated, and therefore broader, while the emission from blazars mostly comes from the strongly boosted, thin spine.
\begin{figure}[!h]
\centering
\includegraphics[trim=0.15cm 0.1cm 0.5cm 0.08cm, clip=true, width=\linewidth]{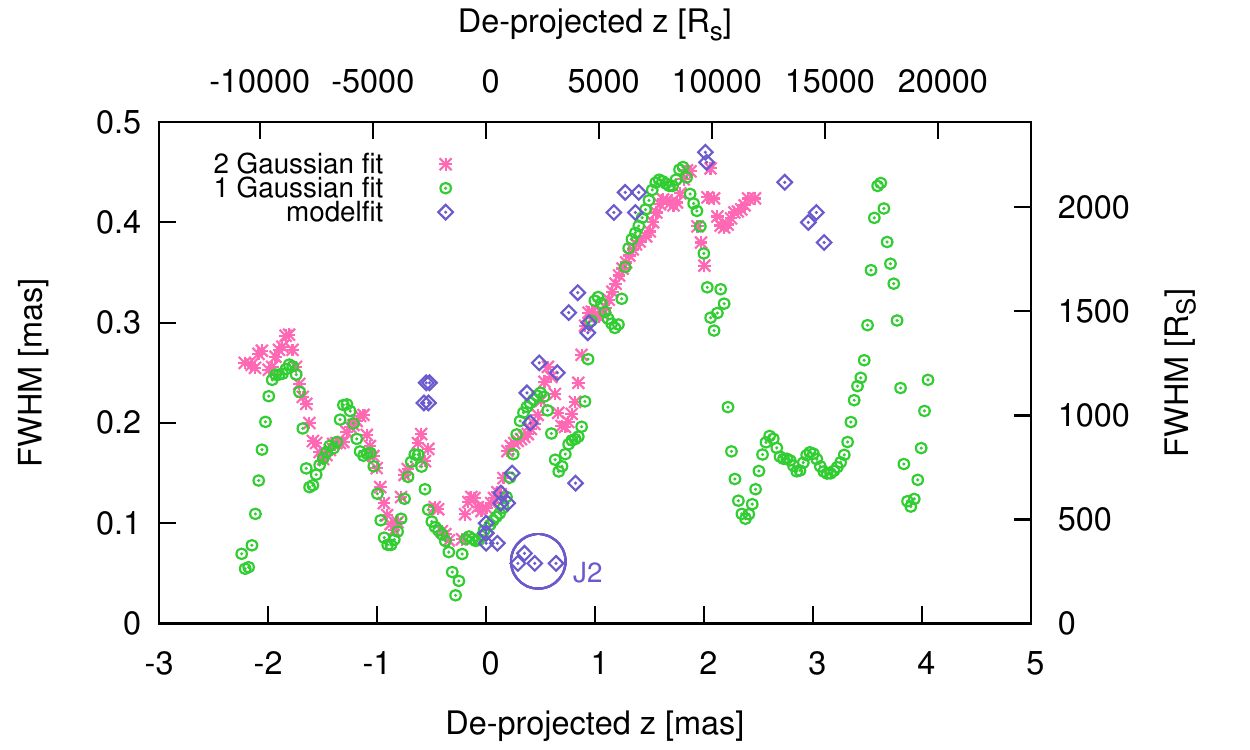}
\caption{\footnotesize{Jet aperture versus core separation in November 2009 from the double ridge line analysis (magenta), from a single Gaussian fit (green). Overplotted in blue are the sizes of the modelfit components from all epochs. Component J2 occupies only a portion of the jet in the transverse direction. Note that, in order to more clearly show the expansion, the $x$ and $y$ axis have different scales.}}
\end{figure}
\subsection{Narrowing of the jet at $z{\sim}2$ mas}
The increase of the jet width is not always smooth, but is characterized by some oscillations. Although an oscillating expansion can occur in jets  \citep{1988ApJ...334..539D}, we cannot exclude that this is an image artifact, especially in the faint regions of the counter-jet. The narrowing at ${\sim}$ 2 mas in the main jet, and perhaps in the counter-jet too is, however, very prominent and coincides with the location of the stationary feature J6. Stationary features in jets can be related to instability patterns, as mentioned in Sect. 4.2, or may result from re-collimation mechanisms. In the purely hydrodynamic case, recollimations naturally occur when the expanding flow becomes under-pressured compared to the medium \citep{1988ApJ...334..539D}. In MHD flows, collimation features can also appear when the jet ceases to be strongly magnetized and causally connected to the central engine \citep{2012bhae.book.....M}. The transition to the new regime is possibly accompanied by a saturation of the acceleration, 
which is nevertheless not observed in this region. Kinematic VLBI studies at lower frequencies \citep[][and references therein]{2003enig.conf..216B} have shown that the acceleration in the main jet extends at least to ${\sim}$5 mas, where another prominent narrowing appears \citep{1998A&A...329..873K}. The nature of this collimation feature is ultimately not clear. It is interesting though that the narrowing is not present if we look at the modelfit widths (Fig. 7). In other words, the larger beam used in the modelfit analysis does not ``see'' the feature, indicating that whatever process is taking place, it must involve the flow on smaller scales. 
\subsection{Collimation regime}
Theoretical models and simulations describing the conversion from Poynting flux to kinetic flux in magnetically-driven jets have shown that the characteristics of the external medium represent a fundamental parameter \citep{2007MNRAS.380...51K,2008MNRAS.388..551T,2009ApJ...698.1570L}. Specifically, the gradient of the external pressure along the jet largely determines the shape of the jet and the acceleration rate, which are intimately connected. In an analytical study of the asymptotic structure of magnetically-driven flows, \cite{2009ApJ...698.1570L} predicts the existence of two main regimes. In the first, the equilibrium regime, the pressure of the external medium $p$ decreases slowly, as $p\propto z^{-k}$ with $k<2$. In these conditions the magnetic field  in the flow succeeds in keeping a strong poloidal component. The jet radius $r$ as a function of distance from the core is then given by a power law with index smaller than 0.5, the Lorentz factor increases as the radius ($\Gamma \propto r$) and the 
asymptotic shape of the jet is cylindrical. If the pressure gradient is steeper, with 
index $k>2$, the poloidal magnetic field becomes negligible, while the toroidal component cannot provide strong confinement to the flow, which then gets broader, accelerates more slowly ($\Gamma \propto z/r$), and eventually reaches a conical shape. This is the non-equilibrium regime. 
The transition state between the two regimes occurs for $k=2$. In this case, the asymptotic behavior of the flow is determined by the intensity of the magnetic field $B_0$ and by the external pressure $p_\mathrm{0}$ at the base of the flow. Specifically, if the ratio $(6\pi p_\mathrm{0})/B_\mathrm{0}^2$ is larger than 0.25, the flow has characteristics similar to the first regime, and has a perfectly parabolic shape, i.e. with a radius $r$ increasing as $z^{0.5}$. A smaller ratio instead leads the flow to a state similar to the second regime. 

In the following, we show that Cygnus A resembles the properties expected in the transition state ($p\propto z^{-2}$) and that the asymptotic regime is that of equilibrium. Let us consider the shape of the jet. Because of the higher SNR, we focus our analysis on the approaching side. In order to reduce the impact of small scale oscillations and to study the overall expansion of the jet, we consider the modelfit sizes in the four epochs up to the narrowing at ${\sim}$ 2 mas. We exclude component J2 because, as shown in Fig. 7, this is clearly a feature occupying, transversally, only a portion of the jet. Results are shown in Fig. 8. The expansion can be described by a single power law of the form $r\propto z^{0.55\pm 0.07}$ up to ${\sim} 10^4 R_\mathrm{S}$, i.e. the jet has a parabolic shape. Two elements suggest that the flow collimates in the equilibrium regime. Firstly, the Lorentz factor gradient of the fast component calculated in Sect. 5.1 ($\Gamma \propto z^{0.50\pm0.19}$) is compatible with the 
prediction that $\Gamma$ grows as $r$, at least in the central body of the jet. Secondly, VLBI studies at 5 GHz \citep{1991AJ....102.1691C} showed that the jet has a cylindrical shape between 4 and 20 mas, which is the asymptotic state in the equilibrium regime. Interestingly, 4--5 mas (${{\sim}}2.5 \times 10^4 R_\mathrm{S}$) is also the maximum scale on which acceleration has been observed in this source \citep{2003enig.conf..216B}, and where a second, prominent recollimation occurs \citep{1998A&A...329..873K}.  Therefore this location can be identified as a good candidate for a region where the jet of Cygnus A ceases to be strongly magnetized and becomes causally disconnected from the central engine. 
\begin{figure}[!h]
\centering
\includegraphics[trim=0.15cm 0.1cm 0.5cm 0.1cm, clip=true, width=\linewidth]{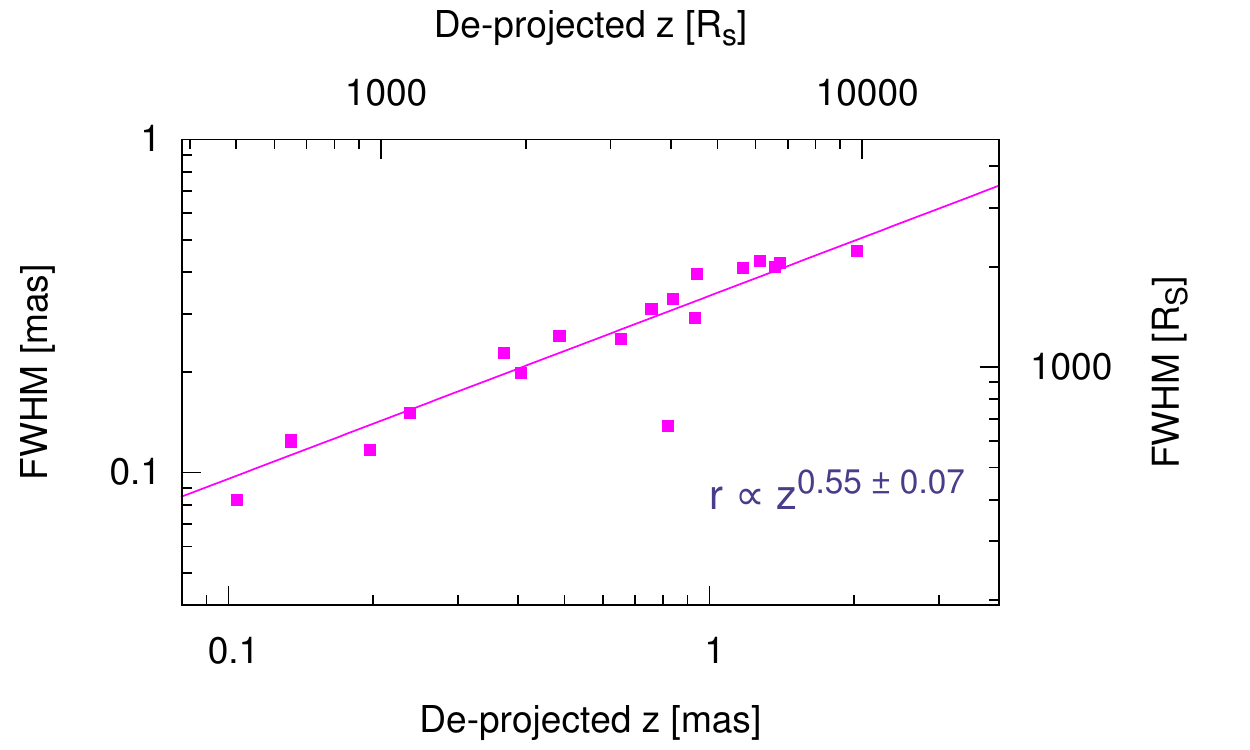}
\caption{\footnotesize{Power-law fit of the jet width (FWHM=2*radius $r$) versus de-projected distance from the core, as inferred from \textsc{MODELFIT}.}}
\end{figure}

\subsection{Comparison with M87}
In the previous section we have shown that the prototypical FRII radio galaxy Cygnus A is collimated and accelerated on a scale of ${{\sim}}2.5 \times 10^4 R_\mathrm{S}$, with the jet being parabolic in this region and cylindrical further downstream. It is interesting to compare these properties with those inferred for M87, the most studied FRI object. \cite{2012ApJ...745L..28A} have shown that the jet of M87 is also parabolic up to a scale of ${\sim} 10^5 R_\mathrm{S}$, with $r\propto z^{0.59}$, thus with a power law index very close to that we measure for Cygnus A. However, after the HST-1 feature, the jet becomes conical rather than cylindrical, just as expected in the non-equilibrium regime. An important difference between the equilibrium and non-equilibrium regimes resides in the efficiency of conversion of Poynting flux. While in the former case the Poynting flux will be mostly converted into kinetic flux, until equipartition is reached, a jet in non-equilibrium will remain 
magnetized up to large distances without reaching very high Lorentz factors. The high power, thin jets of the FRII sources and the lower power, broader jets of the FRIs appear to fit well into the two categories respectively, indicating that the FRI/FRII dichotomy might result from different environmental conditions. The discovery of sources with hybrid morphologies \citep{2002NewAR..46..357G}, the so called HYMORS, supports the same scenario.

\subsection{Nature of the gap of emission}
Finally, we comment on the nature of the sharp gap of emission situated at ${\sim}$ 0.2 mas east of component N and of
${\sim}$ 0.1 mas in size. Figure 7 shows that, although the zero in the $x$ axis is set to coincide with the location of component N, this is not the most narrow region of the jet. Instead, both jet and counter-jet appear to emanate exactly from the gap of emission, which features the absolute minimum width. It is then likely that the gap marks the true location of the central engine, which would then be located ${\sim}$800 $R_S$ upstream of the jet base, represented by component N.
Given the amount of evidence for the jet being magnetically driven, it is interesting to wonder how the jet would appear when it is still completely magnetically dominated, i.e. with a magnetization parameter much larger than 1. Due to the very low speed the flow would have, it would also be faint and may appear as a gap. Other explanations are of course possible. The gap could alternatively be the effect of strong absorption by a molecular torus, which is certainly present and obscures part of the counter-jet \citep{2010A&A...513A..10S}. Another possibility is that this is simply a self-absorbed region. A detailed spectral analysis using higher frequency 
data is required and in progress.

\section{Conclusions}
We have performed a high resolution VLBI study at 7 mm of the kinematic properties and transverse structure of the two-sided jet in Cygnus A. Results can be summarized as follows.
\begin{itemize}
\item We detect an acceleration of the flow up to superluminal speeds ($\beta_{\mathrm{app}}^{\mathrm{max}}=1.15\pm0.04$) in the first parsec of the jet. 
The outer jet exhibits lower speeds, in agreement with studies at lower frequencies. We suggest that a fast and a slow component exist in the flow, and interpret the slower speeds in the outer jet as an effect from the de-boosting of the fast component, naturally occurring at large viewing angle.
\vspace{0.2cm}
\item The acceleration region of both the fast and slow component is extended (${\sim}10^3-10^4 R_\mathrm{S}$), as characteristically predicted for magnetically-driven outflows. 
\vspace{0.2cm}
\item  Both jet and counter-jet are limb brightened. In the presence of a transverse gradient of the bulk Lorentz factor, the limb brightening is evidence for a spine-sheath structure of the jet, as any fast flow appears faint at large viewing angle.
\vspace{0.2cm}
\item The mean opening angles in the two sides are large 
(the full angle in the jet is ${\sim} 10^{\circ}$) compared to typical values inferred for blazars ($1^{\circ}-2^{\circ}$). This can be explained by the possibility, in radio galaxies, to observe not only the thin inner spine but also the outer sheath.\vspace{0.2cm}
\item The shape of the flow is parabolic ($r\propto z^{0.55\pm 0.07}$). The acceleration gradients and the collimation regime can be well reproduced by assuming an external pressure $p$ which scales with distance from the central engine $z$ as $p\propto z^{-2}$, and the ratio $(6\pi p_\mathrm{0})/B_\mathrm{0}^2$ to be larger than 0.25 at the base of the flow, as described in \cite{2009ApJ...698.1570L}. 
\vspace{0.2cm}
\item The jet stops being strongly magnetized possibly on a scale of ${\sim} 2.5 \times 10^4 R_\mathrm{S}$, where it enters the asymptotic cylindrical shape \citep{1991AJ....102.1691C}, the bulk of the acceleration saturates \citep{2003enig.conf..216B} and a recollimation occurs \citep{1998A&A...329..873K}.
\end{itemize}
\begin{acknowledgements} 
B.B. is a member of the International Max Planck Research School (IMPRS) for Astronomy and Astrophysics at the Universities of Bonn and Cologne. B.B. thanks Andrei Lobanov for substantially triggering the ideas of this paper, and Richard Porcas for reading the manuscript and for his constant help. The author is also greatful to Vassilis Karamanavis, Gabriele Bruni, Tuomas Savolainen, Manel Perucho and Robert Laing for stimulating discussions. E.R. acknowledges partial support by the MINECO grant AYA-2012-38491-C02-01 and Generalitat Valenciana grant PROMETEOII/2014/057. The European VLBI Network is a joint facility of European, Chinese, South African and other radio astronomy institutes funded by their national research councils. The National Radio Astronomy Observatory is a facility of the National Science Foundation operated under cooperative agreement by Associated Universities, Inc. 
\end{acknowledgements}
\bibliographystyle{aa}
\bibliography{reference.bib}
\end{document}